\def\kmskpc{\,{\rm km}\,{\rm s}^{-1}\,{\rm kpc}^{-1}}
\def\kms{\,{\rm km}\,{\rm s}^{-1}}
\def\masyr{\,{\rm mas}\,{\rm yr}^{-1}}
\def\kpc{\,{\rm kpc}}
\def\mas{\,{\rm mas}}
\def\spose*1{\hbox to 0pt{*1\hss}}
\def\lta{\mathrel{\spose{\lower 3pt\hbox{$\mathchar"218$}}
     \raise 2.0pt\hbox{$\mathchar"13C$}}}
\def\gta{\mathrel{\spose{\lower 3pt\hbox{$\mathchar"218$}}
     \raise 2.0pt\hbox{$\mathchar"13E$}}}
\def\ex*1{\langle*1\rangle}

\documentclass{aa}

\input epsf

\usepackage{latexsym}

\begin{document}

\thesaurus{04(10.11.1; 10.19.3)}

\title{The Galactic warp in OB stars from Hipparcos} 

\author{ R. Drimmel
	\and R.L. Smart \and M.G. Lattanzi }
\institute{ Osservatorio Astronomico di Torino, Strada
	       Osservatorio 20, I-10025 Pino Torinese, Italy.}

\offprints{drimmel@to.astro.it}
\date{Received: /Accepted:}

\maketitle

\begin{abstract}
The kinematics of distant OB stars perpendicular to the Galactic
plane, inferred from proper motions in the Hipparcos catalogue, are
analysed and compared to the kinematic signature that would be induced
by a long-lived Galactic warp. Previously we reported that the
kinematics of the OB stars toward the anticenter were inconsistent
with the assumption of a long-lived warp (\cite{SMA98}), showing
negative systematic motions as opposed to the expected positive
motions.  Using a larger sample of OB stars, improved distances, and a
more accurate model for the uncertainties, we confirm our previous
result for a larger range of galactocentric radii.

However, we note that the new model for errors in the photometric
distances reveal an important bias that causes the observed systematic
vertical motions to be smaller than their true values.  Using
synthetic catalogues we investigate the effect of this bias on the
systematic vertical motions in conjunction with the possibility of a
smaller warp amplitude and a warp precession rate, both of which can
also lead to smaller systematic motions. Taken together these three
effects can work to produce negative observed systematic vertical
motions, similar to those detected in the data, though only with both
excessively high precession rates ($-25 \kmskpc$) and very large
photometric errors (1 magnitude).

\keywords{Galaxy: kinematics and dynamics -- Galaxy: structure}
\end{abstract}

\section{Introduction}

For some time now the existence of a warp in our Galaxy has been well
known.  First seen in neutral hydrogen emission, it's presence in
several other components have since been reported, i.e. IRAS point
sources (\cite{DJO89}), CO gas (\cite{WOU90}), and dust (\cite{FRE94}).
Studies of the Galactic warp have been mostly limited to it's spatial
structure (see Binney 1992 for review) because the kinematic
signature of a warp primarily manifests itself in the component
tangential to the line-of-sight (LOS) of an observer near the Galactic
plane, the component that cannot be directly measured by radio
observations. For stars, on the other hand, this component is in
principle accessible via proper motions, but in practice this approach
has been limited by the small magnitude of the expected systematic motion 
($\sim1$--2 $\masyr$), which is
smaller than the zonal systematic errors in traditional
catalogues. The advent of astrometry from space, inaugurated with the
Hipparcos satellite, signals a new era in the study of Galactic
kinematics; the high precision ($\leq 1\masyr$) and accuracy 
($<.1\masyr$) of the proper motions combined with and
inertial frame means that it is now possible to detect the kinematic
signature expected from a Galactic warp.  Once the spatial structure
of the warp is determined, its kinematic signature will yield direct
information on the precession rate of the warp.

Earlier we reported results from a subsample of the Hipparcos catalogue
(\cite{SMA98}). In that study we looked exclusively at distant OB stars
toward the anticenter (between 70 and 290 degrees in galactic
longitude), expecting such stars to trace the motions of the gaseous
component from which they were recently born. Unexpectedly it was
found that their kinematics did not follow the predicted signature of
a long-lived warp, either precessing or not. This led to the
conclusion that either there are additional systematic vertical
motions in the Galaxy's gaseous component, or that the warp is not a
long-lived structure.

In this contribution we use a larger sample of OB stars from the
Hipparcos catalogue to reinvestigate their systematic vertical
motions.  As before, our interpretation of the data is based upon a
comparison between the observed kinematics of the selected sample of
OB stars, and the ``observe'' kinematics of a sample created using a
synthetic catalogue generator. The synthetic catalogues are generated
based on an assumed statistical model of the (warped) stellar
distribution, kinematics and observational errors. By making the
identical selections and cuts on the simulated catalogues as are made
on the actual data, any biases present or introduced in the sample of OB
stars due to errors or selection criterion are likewise reproduced in
the synthetic samples as well.  In this way the modeled kinematics are
translated into observed quantities which can be directly compared
with those of the selected sample of stars. In essence, rather than
attempting to remove or correct biases in the observed sample, we
introduce the same biases into the model in order to determine what is
actually detectable.

In the following section the new sample of OB stars is described,
while in Sect. 3 
the model for the errors used for the synthetic catalogues is presented.
Sect. 4 details an improved estimate of the distances and
in Sect. 5, using a simple model for the Galactic warp,
we find spatial parameters consistent the distribution of the Hipparcos
OB stars. In Sect. 6 we discuss
the estimation of the systematic vertical motions, derive the kinematic 
signature resulting from a long-lived precessing warp, and compare the 
observed and expected kinematic signatures. Sect. 7
describes the bias introduced in the estimation of the vertical motions
by errors in the distance and how these, in conjunction with a smaller
amplitude and precession, can produce smaller systematic motions than 
otherwise expected. We present our conclusions in Sect. 8.

\section{The Sample}

The sample of OB stars used here to study 
the Galactic warp is taken from the 10544 OB stars in the Hipparcos
catalogue; after excluding 3 stars whose spectral typing is no more
specific than ``O'' or ``B'', we find 4538 stars with a parallax
$\pi \leq 2\mas$. Stars with measured parallaxes $\pi < 0$ 
are retained, as their exclusion would unfavorably bias the sample 
with respect to distance.

\begin{figure}[ht]
 \epsfysize=6.0cm
 \epsffile{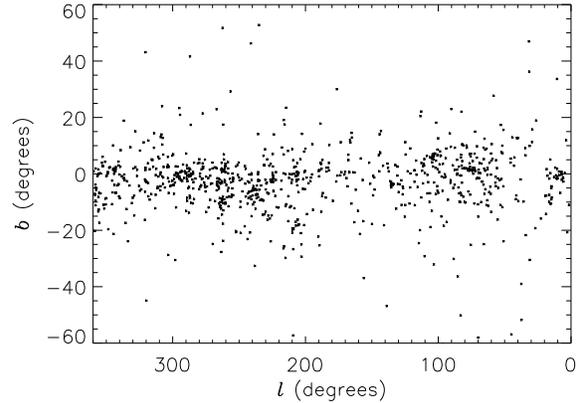}
 \caption{
Distribution of the subsample of 929 distant ($\pi \leq 2\mas$)
Hipparcos OB stars on the sky with apparent magnitudes brighter than 7.5 .
 }
 \label{lbdist}
 \end{figure}

In the Tycho catalogue there are 857 stars brighter than $V$ magnitude 
7.5 which do not appear in the Hipparcos catalogue,  which has a total of 
24384 stars brighter than this magnitude. This suggests that the Hipparcos
catalogue is approximately 97 percent complete to this magnitude limit. 
From our sample alone, based on the distribution of apparent
magnitudes brighter than magnitude 7, we estimate that the subsample of
3840 OB stars with $m \leq 7.5$ is 98 percent complete (\cite{SMA99}). 
Of these brighter stars, 929 have a parallax $\pi \leq 2\mas$.
Fig. \ref{lbdist} shows the distribution of this subsample 
of distant OB stars on the sky.

In our earlier work (\cite{SMA98}) purely photometric distances were used. 
In Sect. 4 we will improve upon these distance estimates, but nevertheless
complete spectral classification is needed to calculate a photometric 
parallax. For the 7279 OB
stars with complete spectral classification supplied by Hipparcos, the
absolute magnitudes and colors from Schmidt-Kaler
(1982) are used to calculate the photometric parallax. 
For 141 stars without luminosity classification H$\beta$
linewidths were used to determine luminosity class. 
The 3121 remaining stars without luminosity classification 
have photometric distances derived from
their reddening and estimated absorptions. 
For further details of these procedures the
reader should consult Smart~et~al. (1997).  

In order to understand the properties of the photometric error, the
sample of all OB stars brighter than 7.5 will be considered in the
following section. These are 3840 in number, with 789 not having an
initial luminosity classification (nonLC stars).  As we expect that
our estimation of the luminosity classes for these stars will be more
susceptible to error than those with full spectral classification, it
is important to be aware of biases in their distribution. Indeed,
because of the use of the Michigan Spectral Survey in the southern sky
(\cite{MSS4} and references
therein), 745 of the bright nonLC stars (94 \%) are in the north ($\delta >
-12.$ deg). This zonal bias results in a photometric error that is
systematically larger in the northern part of the sky than the south,
as the following section describes.

 \begin{figure}[ht]
 \epsfysize=6.0cm
 \epsffile{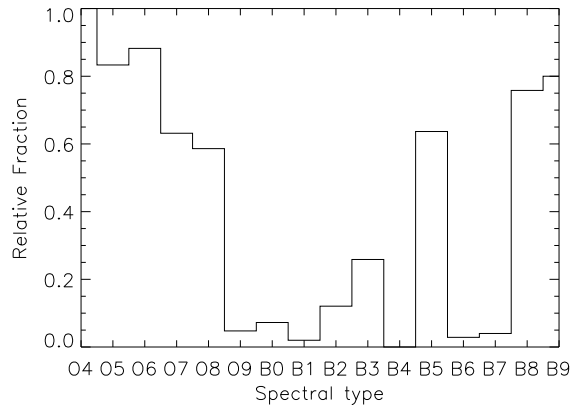}
 \caption{
The distribution of nonLC stars in the north ($\delta > -12.$ deg) with
respect to spectral type.
 }
 \label{nonlcsp}
 \end{figure}

 \begin{figure}[ht]
 \epsfysize=7.0cm
 \epsffile{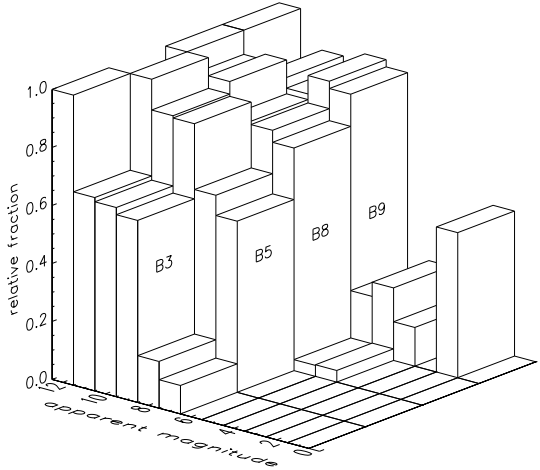}
 \caption{
The distribution of the northern nonLC stars in apparent magnitude
for the spectral types B3, B5, B8 and B9.
 }
 \label{nolclego}
 \end{figure}

To account for the higher photometric error introduced by the nonLC stars, it
is necessary to describe their distribution not only on the sky,
but also with respect to other relevant variables. Fig. \ref{nonlcsp} 
shows the relative fraction of northern nonLC stars with respect to
spectral type and Fig. \ref{nolclego} the relative fraction of 
nonLC stars with respect to
apparent magnitude for B9, B8, B5 and B3 spectral types. The fact that
the nonLC stars are primarily restricted to these types,
and not hardly seen in types B7, B6, and B4, indicates that
the nonLC stars are in general typed at a lower resolution than
those stars with luminosity class. For stars of spectral types earlier 
than B2 the relative fraction of nonLC stars does not vary strongly with
apparent magnitude.

While in Sect. 4 only a magnitude cut is applied, in Sects. 5 and 6
two additional cuts are employed; stars with a parallax of $\pi >
2\mas$ and a vertical velocity greater than 50 $\kms$ are removed.
The parallax cut is imposed to avoid being biased by local structure,
such as Gould's Belt, which dominates the distribution of nearby OB
stars.  The velocity cut is astrophysically motivated by the presence
of runaway stars, which through close encounters have reached escape
velocities, and thus do not possess kinematics principally determined
by the Galactic potential. The typical threshold detection value
for the relative space velocity of runaway stars varies from 40 to 65
$\kms$ (Blaauw~1993,~Torra~et~al.~1997),\nocite{BLA93}\nocite{TOR97}
thus our value of 50 $\kms$ applied to one component is conservative,
especially as it is approximately seven times greater than the
velocity dispersion of this population. The velocity cut is applied
to galactocentric radial bins, excluding those stars which deviate
from each bin's median velocity by more than $50\kms$, reducing the
sample of bright, distant OB stars to 895 in number (less 34), 
142 of which are nonLC stars. Meanwhile 255 are cut from all the
stars with $\pi \leq 2$, leaving 4283 stars, 1291 of which are nonLC
stars. The calculation of the individual and mean vertical velocity is
discussed in Sect. 6.

\section{Simulated catalogues and improved error model}

In order to interpret the observed kinematics of OB stars from the
Hipparcos catalogue, a program was developed to generate synthetic
catalogues based on a statistical model of the spatial 
and velocity distributions of the OB stars (Drimmel~et~al. 1997, 1999). 
The model distribution is non-axisymmetric, being described by spiral arm
segments, most important of which is the local Orion arm, which has 
a pitch angle of 8 degrees and a spur projecting toward the outer Perseus
spiral arm. The synthetic catalogue generator produces catalogues
with distributions in galactic longitude and latitude ($l,b$), 
apparent magnitude, and proper motion, which closely resemble those
of the distant ($\pi \leq 2\mas$) OB stars in the Hipparcos catalogue
(see also \cite{SMA97A}). 

The synthetic catalogue generator uses a Galactic dust distribution model 
based on the 240$\mu$ COBE data (Drimmel \& Spergel, in preparation)
to calculate absorbtions. This model is
a refined version of the earlier work of Spergel~et~al. (1997), adding 
nonaxisymmetric structure to the dust distribution model,
the most important of which is dust associated with the local Orion arm. 
The added structure in the dust distribution results in additional reddening 
in the directions tangent to the local arm, thus improving the modeled
longitude distributions (see Drimmel~et~al. 1999 as compared to 
Drimmel~et~al. 1997). This same model is used to estimate reddenings and 
absorptions to the nonLC stars.

Observational errors are modeled for generating observed quantities.
The simulated catalogues in Smart~et~al. (1998) used
parallax errors, $\sigma_{\rm t}$, equal to 1\,mas, and proper motion errors, 
$\sigma_{\mu}$, were described
by simple functions of galactic latitude, as seen in the Hipparcos 
catalogue (\cite{mignard97}). These errors are actually
better described as varying with ecliptic latitude and apparent
magnitude. We have thus improved our model for 
$\sigma_{\rm t}$ and $\sigma_{\mu}$ 
by interpolating from Tables 3.2.4-6 given in the first volume of the
Hipparcos catalogue (\cite{HIPCAT}). 
We have also added a 0.01 magnitude error to the apparent magnitudes. 
For our purposes here the Hipparcos errors in position ($l,b$)
are negligible. 

\begin{figure*}[ht]
\epsffile{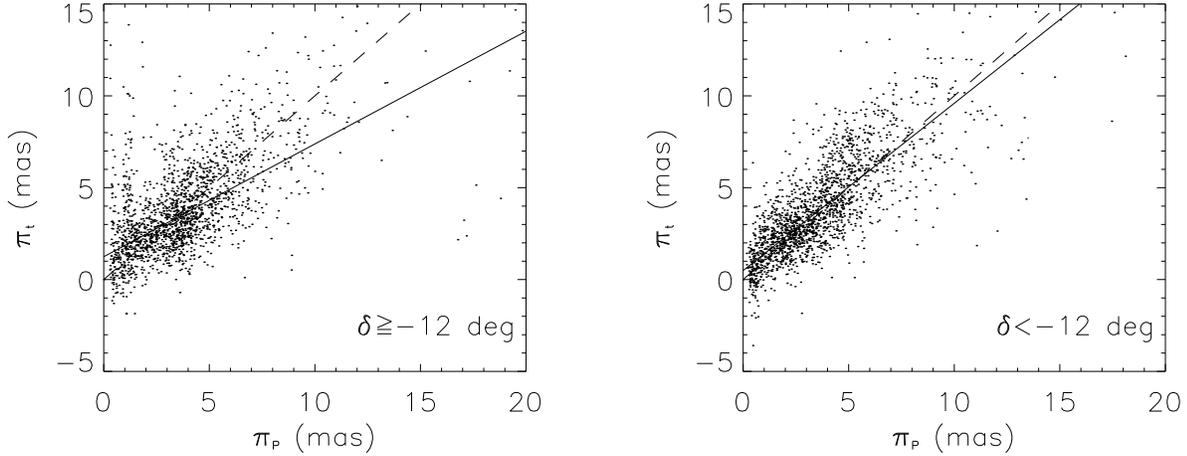}
\caption{
Photometric versus trigonometric parallax distributions of the bright
($m < 7.5$) Hipparcos OB stars for the north ($\delta > -12.$ deg) and 
south. The dashed line has a slope of 1, given as reference, while
the solid line is the result of a robust linear fit that minimizes 
the absolute deviations (no points excluded). 
}
\label{pipi_all}
\end{figure*}

\begin{figure*}[ht]
\epsffile{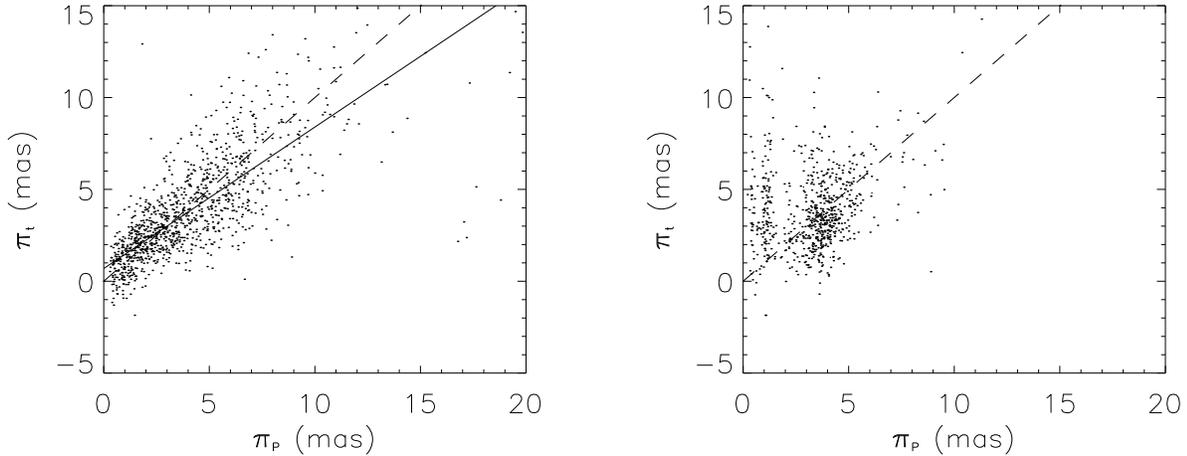}
\caption{
The photometric-trigonometric parallax distributions of the 
bright Hipparcos OB stars in the north, decomposed into stars
with initial luminosity classes (left plot) and without
initial luminosity classes (right plot). 
Dashed and solid lines are as in the previous figure.
}
\label{pipi_nord}
\end{figure*}

However, the improvement of the model of
photometric distance error makes the greatest difference in the
simulated observable kinematics of the synthetic catalogues.
Previously our error model involved adding a relative gaussian error 
of 25 percent of the distance.
However, if the errors are normally distributed in absolute magnitude,
it is the errors in the distance modulus that are normally distributed.
We now model the error in photometric distance by assuming a normal distribution
of errors in the absolute magnitude.
This error is not only the result of the intrinsic scatter 
of absolute magnitudes, termed ``cosmic error'', but also due to errors 
introduced by uncertainties in the determination of the absorption, intrinsic
scatter in the colors, and misclassification errors. Not all these 
errors are guaranteed to be gaussian.

The photometric error of our observed sample of stars is evaluated using 
the photometric parallax $\pi_{\rm P}$ 
verses trigonometric parallax $\pi_{\rm t}$
distribution for the sample of 3840 OB stars brighter than 7.5. 
Fig. \ref{pipi_all} shows the $\pi_{\rm P}$--$\pi_{\rm t}$ 
distributions for the north
and the south. The degraded photometric parallaxes of the north, due
to the presence nonLC stars, are 
responsible for the larger scatter as compared to the south. To
illustrate this effect Fig. \ref{pipi_nord} shows the 
$\pi_{\rm P}$--$\pi_{\rm t}$
diagrams for the north decomposed into the stars with and without 
luminosity class. It can be seen that the northern stars which 
possess full initial classification (Fig. 5 left) have approximately 
the same scatter as 
the southern stars (Fig. 4 right), and we will assume the same error model can
be used for both. One difference between the northern and southern stars 
with full initial classification is a systematic bias of $\pi_{\rm P}$ in
the south for stars with $\pi_{\rm t}$ centered at 7.5\,mas.
This bias is due to the presence of stars in the direction of $\rho$ Ophiuchus
whose absorptions are systematically overestimated.

\begin{figure*}[ht]
\epsffile{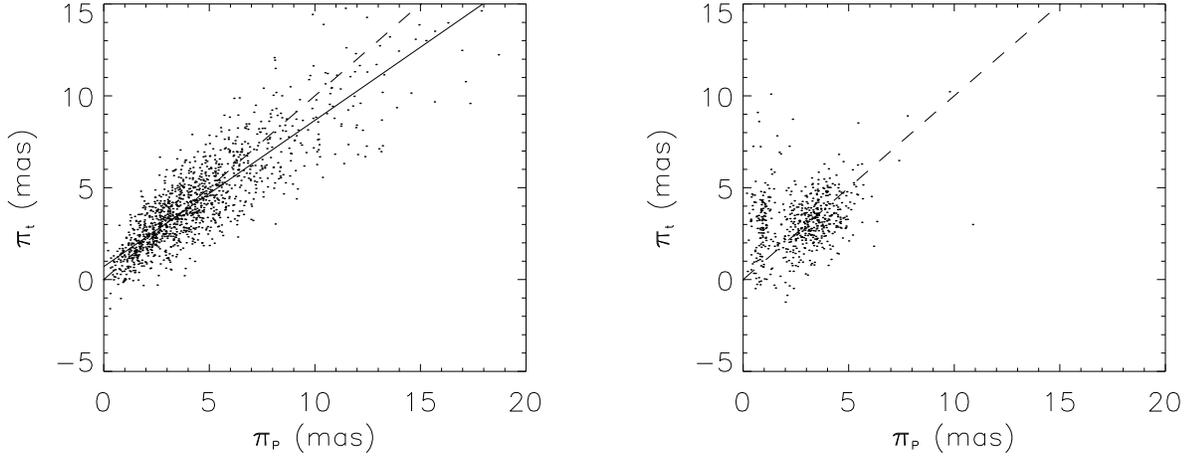}
\caption{
Modeled photometric-trigonometric parallax distribution of bright
OB stars in the north, for LC and nonLC stars (left and right 
plots respectively), as produced by the synthetic catalogue
generator. 
Dashed and solid lines are as in the previous figures.
}
\label{pipi_mod}
\end{figure*}

\begin{figure}[ht]
 \epsffile{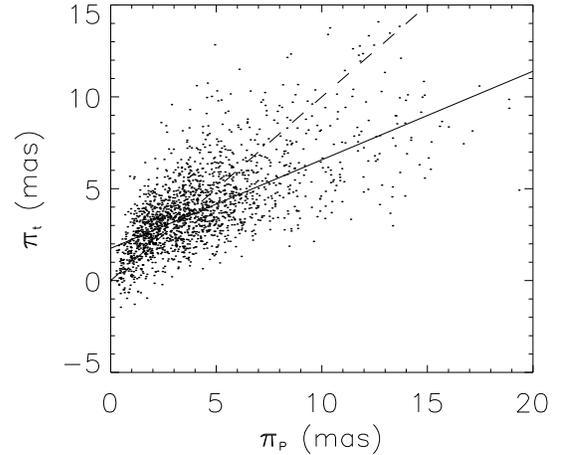}
 \caption{
Modeled photometric-trigonometric distribution of bright OB stars in the 
north assuming $\sigma_M = 1$ magnitude. Dashed and solid lines are 
as in previous figures.
 }
 \label{sigm_1}
 \end{figure}

Assuming that the observational errors in the Hipparcos trigonometric 
parallax are well 
described, we can estimate the photometric error using the synthetic
catalogue generator, comparing observed and generated 
$\pi_{\rm P}$--$\pi_{\rm t}$ distributions. 
For the northern stars with luminosity classification an
error model which approximately reproduces the observed distribution of
trigonometric and photometric parallaxes is 
\begin{equation}
\sigma_M = .5 + {\rm max}(0,(m-6.5)/6) - M/12,
\label{photerr}
\end{equation}
where the apparent magnitude term  
can be interpreted as being the result of a higher 
occurrence of classification error at fainter magnitudes, and the absolute 
magnitude term can be interpreted as an increasing cosmic error
and/or uncertainty for the more luminous stars. 
To characterize the distribution a robust linear fit is made
to the observed $\pi_{\rm P}$--$\pi_{\rm t}$ distribution in the north
(Fig. \ref{pipi_nord}), giving a slope of 0.77 and a mean
absolute deviation in $\pi_{\rm t}$ of 1.28, after 17 stars are excluded
whose absolute deviations in $\pi_{\rm t}$
are greater than 6 mean absolute deviations (presumably misclassified
stars). Meanwhile the modeled distribution (Fig. \ref{pipi_mod}) 
produces a slope of 0.80, with 1.06 being the mean absolute deviation.
For comparison, Fig. \ref{sigm_1} shows the 
$\pi_{\rm P}$--$\pi_{\rm t}$ distribution
for a model generated with a purely gaussian magnitude error of 
$\sigma_M = 1$ magnitude, showing that such a high error does not 
reproduce well the observed $\pi_{\rm P}$--$\pi_{\rm t}$ distribution.

For the nonLC stars the signatures due to misclassification are prominent, 
its nongaussian nature leaving clear artifacts in the distribution,
especially at $\pi_{\rm t} > 2\mas$ and $\pi_{\rm P} < 3\mas$ 
(Fig. \ref{pipi_nord} right). 
These are intrinsically dim stars incorrectly assigned a high luminosity. 
In contrast, intrinsically bright stars misclassified as low luminosity stars 
do not produce clear signatures, but do cause stars of low $\pi_{\rm t}$ to
be systematically biased toward higher $\pi_{\rm P}$. We model the
misclassification errors as offsets to the true absolute magnitudes,
with a binned gaussian probability of misclassification. 

 \begin{figure}[ht]
 \epsfysize=7.0cm
 \epsffile{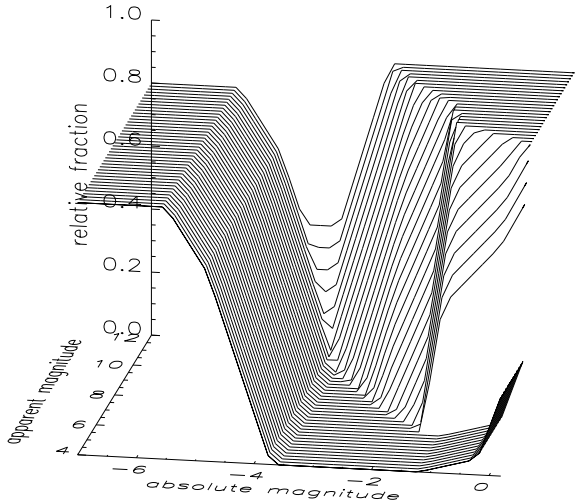}
 \caption{
Model of the relative fraction of nonLC stars in the northern 
part of the sky, as used in the synthetic catalogue generator. 
 }
 \label{nonlcmod}
 \end{figure}

However, before applying such errors the distribution of nonLC stars 
must be modeled. The synthetic catalogue generator does not at this
time generate colors, so spectral types were simply correlated with
the main sequence absolute magnitudes to infer a model of the 
distribution of nonLC stars in the north. Fig. \ref{nonlcmod} shows 
the model of the relative fraction of nonLC stars in the north. 
Once a star is randomly determined to be nonLC, a binning error
is calculated using
\begin{equation}
p={\rm int}(1.5 N + 3 {\rm log} d_0 + 1.4),
\label{offstep}
\end{equation}
where $N$ is a gaussian deviate of unit variance and $d_0$ the true distance, 
$(3 {\rm log} d_0 + 1.4)$ being a bias reflecting that nearby stars are
more likely to be given an over-luminous classification,
and distant stars a under-luminous classification. The resulting
index $p$ is then used to indicate the appropriate offset,
as given in Table \ref{offsets}. In addition to this offset,
gaussian noise of $\sigma_M = 0.3$ magnitude is 
added to the absolute magnitudes.

Much of the added complexity of misclassification in the nonLC stars
will not be needed in what follows, as for stars with $\pi \leq 2\mas$ 
only the positive offsets will contribute significantly to the error.

\section{Improved distances}

In order to improve upon the distances, a distance estimate is used that 
corresponds to the weighted mean parallax of the trigonometric and 
photometric parallaxes:
\begin{equation}
\frac{1}{d} = \pi_{\rm w} = 
\frac{\pi_{\rm t}/\sigma_{\rm t}^2 + \pi_{\rm P}/\sigma_{\rm P}^2}
{1/\sigma_{\rm t}^2 + 1/\sigma_{\rm P}^2},
\label{distance}
\end{equation}
where $\sigma_{\rm P}=\pi_{\rm P} \sigma_M / 2.17$ (\cite{Smith85}), and 
$\sigma_{\rm t}$ is interpolated from Tables 3.2.4-6 in volume one of 
the Hipparcos catalogue. For this distance estimate stars with 
initial full spectral classification
are assumed to have a $\sigma_M$ as specified by Eq. (\ref{photerr}),
while nonLC stars are assigned $\sigma_M=3$ magnitudes. This prescription is
used for both synthetic and observed catalogues.

\begin{table}[ht]
\label{offsets}
\caption{ 
Modeled offsets in absolute magnitude due to misclassifications,
used according to value of the binning error parameter $p$ generated
according to Eq. (\ref{offstep}).
}
\begin{center} 
\leavevmode
\begin{tabular}{llll}
      \hline \\[-5pt]
 $p$ & offset  & Magnitude range & Misclass.  \\
[+5pt]\hline \\[-5pt]
$-$4	&  $-6.8 - M$		&   		& V $\rightarrow$ Ia	\\
$-$3	&  $-5 - 2M/2.45$	&   		& V $\rightarrow$ Ib	\\
$-$2	&  $-1$			& $M > -1.2$    & V $\rightarrow$ III	\\
	&  $-(5.9+M)/2.45$	& $M \leq -1.2$ & III $\rightarrow$ II  \\
$-$1	&  $-3 - M/2.45$	& 		& V $\rightarrow$ II	\\
 1	&  1.			&		& III $\rightarrow$ V 	\\
$>$1	&  1.			& $M > -3.1$    & III $\rightarrow$ V \\
	&  2.5			& $-5.2<M\leq -3.1$ & II $\rightarrow$ V \\
	&  3.5			& $M\leq-5.2$	& I $\rightarrow$ V \\
\hline \\
\end{tabular}
  \end{center}
\end{table} 

\begin{figure*}[ht]
\epsffile{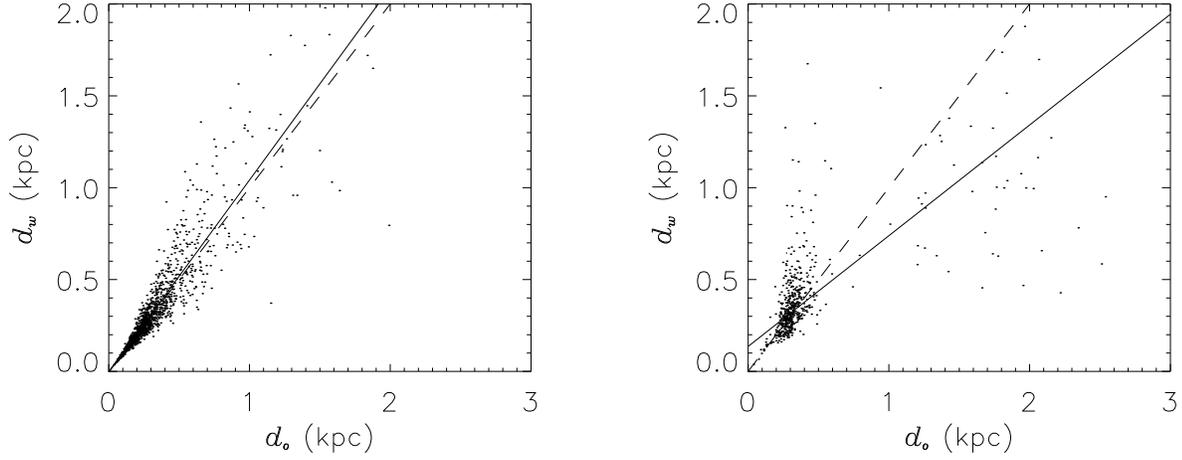}
\caption{
True versus weighted mean distance (Eq. (\ref{distance})) of stars
in a synthetic catalogue for stars with initial luminosity class
(left plot) and nonLC stars (right plot). 
Dashed and solid lines are as in previous plots. 
}
\label{newdist}
\end{figure*}

\begin{figure*}[ht]
\epsffile{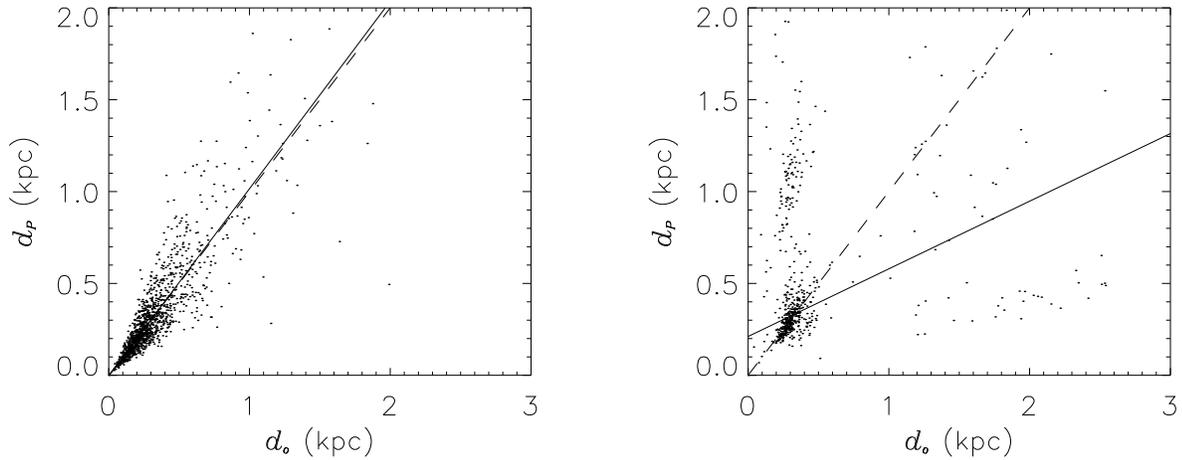}
\caption{
True versus photometric distance of stars
in a synthetic catalogue for stars with initial luminosity class
(left plot), and nonLC stars (right plot) where misclassification
artifacts are evident. 
Dashed and solid lines are as in previous plots. 
}
\label{photdist}
\end{figure*}

To test whether this new distance estimator is more or less biased than
the photometric distance alone, the synthetic catalogue generated in the
previous section for the northern sky is used to compare
the estimated distance with the true distance. Figs. \ref{newdist}
and \ref{photdist} show the result of this Monte Carlo test; 
for stars with an initial luminosity class, the 
variance and skew of the deviations from a robust linear 
fit to the photometric distances, are 2.02 and 1.04, 
while with the new distance estimate they are 0.41 and -0.07
respectively, showing that the uncertainties are effectively
reduced. The slopes and intercepts of both distance estimates show no
significant systematic biases.
For the nonLC stars the new distance estimator reduces the scatter
produced by misclassifications, though the distances can still have
significant error.

Using the typical errors $\sigma_M=.7$ and $\sigma_{\rm t}=1.2\mas$,
the error of $\pi_{\rm w}$ for a star at $\pi_{\rm t}=\pi_{\rm P}=1\mas$ 
is about .3\,mas,
for which the distance estimate would be $1^{+.45}_{-.24}\kpc$. 
We also point out that for stars in our $\pi_{\rm t} \leq 2\mas$ sample,
the photometric distance will dominate in Eq. (\ref{photdist}), as it
is only at a distance of $d = \sigma_M/(2.17\sigma_{\rm t})$ that
$\sigma_{\rm P} = \sigma_{\rm t}$.

\section{Warped spatial distribution}

\begin{table*}[ht]
\caption{ 
Angular tilts, $\theta$, of galactocentric rings, in units of degrees, 
implied by the distribution of bright ($m \leq 7.5$), 
distant Hipparcos OB stars in heliocentric $z$ and $y$ coordinates. 
($\theta = \arctan{b}$, where $z=a+by$)
For the warped and non-warped cases (columns 4, 5, and 6) the tilts and
uncertainties are the means and standard deviations of 30 simulated 
catalogues. The warped cases (columns 5 and 6) were generated with 
$r_{\rm w} = 6.5\kpc$ and $r_{\rm h} =$ 15 and 30\,kpc respectively (see Eq.
(\ref{zwarp})). Column 2 gives the number of stars in each bin for
the data; the 92 stars
not appearing in the table are outside the range of the bins.
}
\begin{center} 
\leavevmode
\begin{tabular}{llllll}
      \hline \\[-5pt]
Bin (kpc)     &  N  &	Data  &     No warp       &   Warp          
	&  $1/2$ Warp \\[+5pt]\hline \\[-5pt]
6.5$< r <$7.5 & 140  & $-$0.39 & $-$0.24 $\pm$ 0.94 &  0.34 $\pm$ 0.91 
	&  0.29 $\pm$ 0.73 \\
7.5$< r <$8.5 & 500 &  1.67 &  $-$0.34 $\pm$ 0.44 &  1.08 $\pm$ 0.49 
	& 0.25 $\pm$ 0.44 \\
8.5$< r <$9.5 & 161 &  2.05 &   0.07 $\pm$ 0.93 &  2.54 $\pm$ 0.86 
	&  1.45 $\pm$ 0.80 \\
\hline \\
\label{slopes}
\end{tabular}
  \end{center}
\end{table*} 

As the Sun is near the line of nodes of the Galactic warp the vertical 
deviation from the Galactic plane is small, but a local slope with 
respect to the Galactic plane is produced in the stellar distribution. 
Using the subsample of the 894 bright ($m \leq 7.5$ magnitudes), distant 
($\pi_{\rm t} < 2\mas$) OB stars,
the slopes of the stellar distribution are found in the 
heliocentric coordinates $z$ and $y$, $y$ being 
taken in the direction of rotation, for bins in galactocentric radius $r$, 
($r_\odot \equiv 8\kpc$). The slopes
were found using a standard robust linear fitting routine
which minimizes the absolute deviations in $y$.
In Table \ref{slopes} we express these slopes as a tilt 
angle $\theta$, corresponding
to the tilt of a galactocentric ring whose axis of tilt goes through
the Sun, consistent with the assumption that the line of nodes of the 
Galactic warp goes through the Sun.

It is well known that OB stars are preferentially found in
associations, resulting in a ``clumpiness'' of the distribution on a
larger scale than would be found in a kinematically relaxed
population. Nevertheless, we do not expect the presence of
associations to affect our results above, as we are describing the
overall distribution of OB stars on a scale significantly larger than
the size of a typical association. To test this presumption stars 
which were found
to be within the spatial limits of a given association were identified. For
this purpose the positions ($l,b,d$) and limits 
($\pm \Delta l, \pm \Delta b$) for 38 associations were taken 
from the summary list in Lang (1992), while the 
line-of-sight extent of each association 
($\Delta d = d {\rm max}(\Delta l, \Delta b)$) 
was doubled to take into account possible errors in the distance.
Only four associations have ten or more stars from our
complete subsample, these being the Cygnus 7 (10 stars), Carina (13 stars),
Cepheus 2 (19 stars), and Orion (17 stars) associations, where a total of 
102 stars are found inside all associations. Removing each
of these sets of stars in turn, the tilts were redetermined. 
It was found that only one of the associations (Cepheus 2) 
influenced the measured
slopes by more than 0.5 deg. 

A warp is effected in the modeled distribution by describing the vertical
distribution as a function of $z' = z - Z_{\rm w}(r,\phi)$, where the function 
$Z_{\rm w}(r,\phi)$ describes the spatial form of the 
warp in galactocentric cylindrical coordinates,
$\phi$ being taken in the direction of rotation. 
From radio observations the warp is seen to be symmetric out to about
$r = 16\kpc$, and can generally be described by 
\begin{equation}
Z_{\rm w} = h(r) \sin (\phi - \phi_{\rm w} + \omega_{\rm p} t) ,
\label{zwarp}
\end{equation}
$\phi_{\rm w}$ being the phase of the warp, and $\omega_{\rm p}$ the precession rate 
of the warp in the direction {\em opposite} to Galactic rotation, and 
with 
\begin{equation}
h(r)=\left\{\begin{array}{ll}
(r-r_{\rm w})^2/r_{\rm h}, & ~r > r_{\rm w} \\
       0,      	& ~r \le r_{\rm w}
\end{array}
\right. 
\label{height}
\end{equation}
being a height function parameterized by $r_{\rm h}$ and the radius at which 
the warp starts, $r_{\rm w}$. The radio observations  
show that the Sun lies close to the line of nodes, that is, 
$|\phi_{\rm w}| < 10$ degrees. As the systematic velocity is not sensitive to 
$\phi_{\rm w}$ in this range (\cite{SMA96A}) we have assumed $\phi_{\rm w} = 0$ 
in this study. The assumption of a long-lived warp is effectively
implemented by taking $\omega_{\rm p}$ to be a constant with respect to $r$ 
and time $t$.

Table \ref{slopes} shows the mean tilts obtained from thirty synthetic 
catalogues with and without a warp, one warped distribution being generated
with the parameters $r_{\rm h}=15\kpc$ and $r_{\rm w}=6.5\kpc$, while a second model with
half this amplitude ($r_{\rm h}=30\kpc$) is also shown for comparison. 
The first set of parameters produce tilts that are consistent with 
the observed tilts, and these are adopted as our estimated warp
parameters in what follows. As noted in our earlier work,
the observed tilts indicate a Galactic warp starting within the Solar Circle.

The use of a robust linear fitting routine does not provide formal
errors on individual fits, hence no error is given
on the tilts of the data. In the case of the synthetic 
catalogues we have the freedom to directly evaluate the uncertainty
by generating multiple catalogues, as we have done. Inasmuch as
the synthetic catalogues are accurate statistical representations 
of the observed data, the uncertainties of their tilts will be 
representative of the uncertainties in the tilts of the 
observed catalogue. However, the synthetic catalogues do not 
include the clumpiness due to the presence of OB associations,
which will increase the uncertainties. The 0.5 deg influence of 
the Cepheus 2 association on the tilts can be taken as an
estimate of the probable additional random error that should 
be added to the uncertainties quoted in Table \ref{slopes}.
In any case, the observed tilts are significant and inconsistent 
with the no-warp hypothesis.

Having estimated the parameters that describe the spatial form of the warp
in the OB stars, we are now prepared to consider the kinematic signature 
accompanying such a warp.

\section{Kinematic signature of the warp}

The presence of a long-lived warp in the stellar disk will result in
systematic motions perpendicular to the Galactic plane; for an observer in
the plane of the Galaxy such motions will be primarily observable in the
galactic latitude component of the proper motion. 
Taking ${\bf x} = (\cos l \cos b ,\sin l \cos b, \sin b)$ 
as the unit pointing vector to a star, the
unit vector tangential to the line-of-sight (LOS) in the direction of
increasing galactic latitude is
\begin{equation}
{\bf b} = \frac{ {\bf x} \times ({\bf k} \times {\bf x}) }
              {|{\bf x} \times ({\bf k} \times {\bf x})|}
       = \frac{{\bf k} - {\bf x} \sin b}{\cos b},
\label{bvector}
\end{equation}
where ${\bf k} = (0,0,1)$, the unit vector perpendicular to the Galactic
plane. The component of a star's relative motion in the direction of
${\bf b}$ is
\begin{equation}
( {\bf v} - {\bf v_\odot} ) \cdot {\bf b} = 4.74 d \mu_b,
\label{bmotion}
\end{equation}
where $\bf{v_\odot}$ is the solar motion, $\bf{v}$ the star's velocity with
respect to the local standard of rest (LSR), $d$ and $\mu_b$ being the 
distance in kiloparsecs and the galactic latitude proper motion in 
${\rm mas\,yr}^{-1}$ respectively. Taking
${\bf v_\odot} = (U,V,W)_\odot$ and ${\bf v} = (U,V,W)$, defined in the same
right hand coordinate system as $\bf{x}$, one can write from Eqs.
(\ref{bvector}) and (\ref{bmotion}): 
\begin{equation}
W = \frac{4.74 d \mu_b}{\cos b} + W_\odot + (S-S_\odot) \tan b,
\label{zmotion}
\end{equation}
where $S$ is the component of the star's velocity parallel to the Galactic
plane and in the plane that contains the LOS and is at right angles to 
the Galactic plane, that is $S=U \cos l  +  V \sin l$,
and similarly for $S_\odot$. Since distant stars are being considered, we do
not wish to assume that the stars are in the solar neighborhood (Oort's
approximation), and $S$ will
therefore contain a contribution from both differential Galactic rotation
and the star's peculiar motion. Neglecting the peculiar velocity contribution
for the moment, we can write
\begin{equation}
S = \left( \overline{v}_\phi \frac{r_\odot}{r} - v_{\rm LSR} \right) \sin l,
\label{difrot}
\end{equation}
where $\overline{v}_\phi$ is the average rotational velocity of the
stellar population, and $v_{\rm LSR} = 220\kms$.
Because the OB stars do not follow Stromberg's asymmetric drift
equation (\cite{DB98b}), we simply take 
$\overline{v}_\phi = v_{\rm LSR} - 3.0(r-r_\odot)$, 
found to be consistent for this stellar population (\cite{DRI97A}).

\begin{figure*}[ht]
\epsfysize=10.0cm
\epsffile{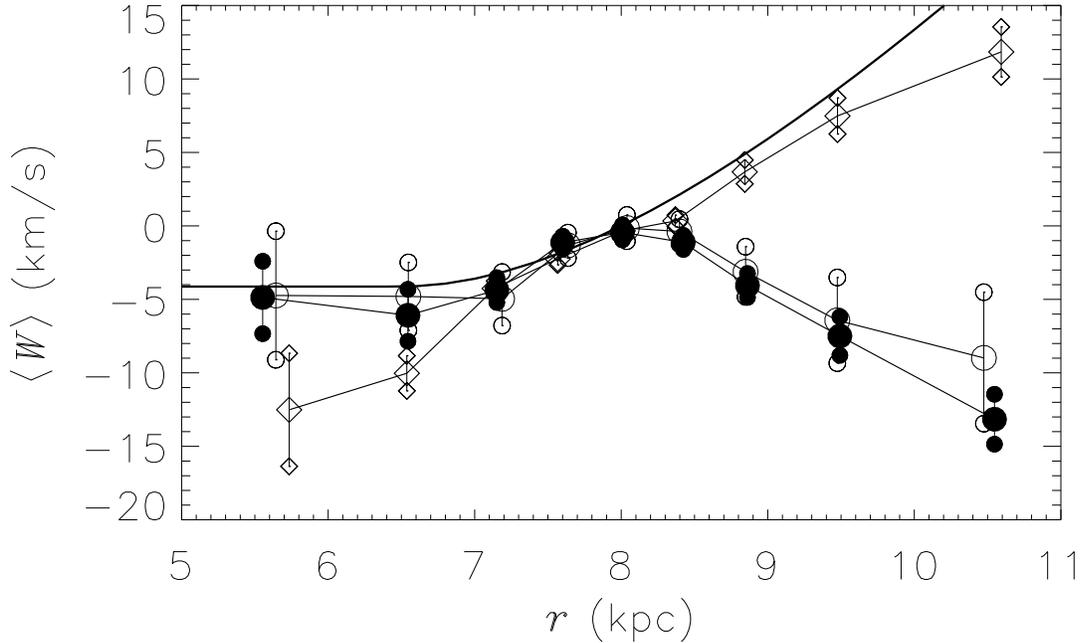}
\caption{The systematic vertical velocity $\langle W \rangle$, relative 
to the LSR, with respect 
to galactocentric radius. Filled circles are for 4250 distant 
Hipparcos OB stars
to 13th magnitude, while open circles show the signal from the complete portion
of our sample (894 stars to magnitude 7.5). 
The thick solid line is $\overline{v}_z(r) - \overline{v}_z(r_\odot)$
(using Eq. (\ref{sysvel})), the
velocity along the line of nodes for a long-lived non-precessing
warp with an amplitude consistent with the observed spatial distribution, 
while the diamonds show the observable
signature from a single simulated catalogue with the same warp parameters.
The symbols show the median $W$ in each radial bin, 
and their error bars are the standard deviation of the mean of $W$. 
The size of the bins increase exponentially from $r_\odot=8\kpc$ with
a incremental factor of e$^{.2}$, with the exception of the first and
last bins that are e$^{.6}$ larger.
}
\label{vzrfig}
\end{figure*}

Finally, adopting $(U,V,W)_\odot = (9,5,7)\kms$ (\cite{DRI97A}),
with positive $U$ being radially inwards, 
we can calculate an observed $W$ for each star using Eq. (\ref{zmotion}). 
(Our adopted value for the Solar Motion is comparable with
other recent determinations (i.e. \cite{DB98b}), but it should be
noted from Eq. (\ref{zmotion}) that $W_\odot$ enters only as an
offset term.)
The stars are binned in galactocentric radius to find the observed systematic 
vertical velocity $\langle W \rangle$ as a function of $r$; 
this is shown in Fig.\ref{vzrfig}. The agreement of the complete portion 
of our sample with the entire sample is consistent with the assumption 
that stars fainter than 7.5 form a kinematically unbiased sample.

By ignoring the peculiar stellar velocities parallel to the 
Galactic plane, $S_*$, in Eq. (\ref{difrot}), we have effectively
ignored a $\langle S_* \tan b \rangle$ contribution in each bin.
This is done from necessity; we cannot calculate $S_*$ for
stars from the Hipparcos catalogue as we do not have their complete 
space motion. However, from Monte Carlo tests we find that 
$\langle S_* \tan b \rangle$ is typically less than $0.05\kms$.

We now compare the observed vertical systematic velocities $\langle W \rangle$
with those expected from a long-lived warp. As the vertical velocity 
$W$ of each star is with respect to the LSR,
$W = W_* + \overline{v}_z(r,\phi) - \overline{v}_z(r_\odot,0)$,
$W_*$ being the star's peculiar vertical motion, and
$\overline{v}_z(r,\phi)$ the systematic vertical velocity at $(r,\phi)$
resulting from the warp. Upon averaging
$\langle W \rangle = \overline{v}_z(r,\phi) - \overline{v}_z(r_\odot,0)$,
the difference in the systematic vertical motion at the star's and the
Sun's position. 

\begin{figure*}[ht]
\epsfysize=10.0cm
\epsffile{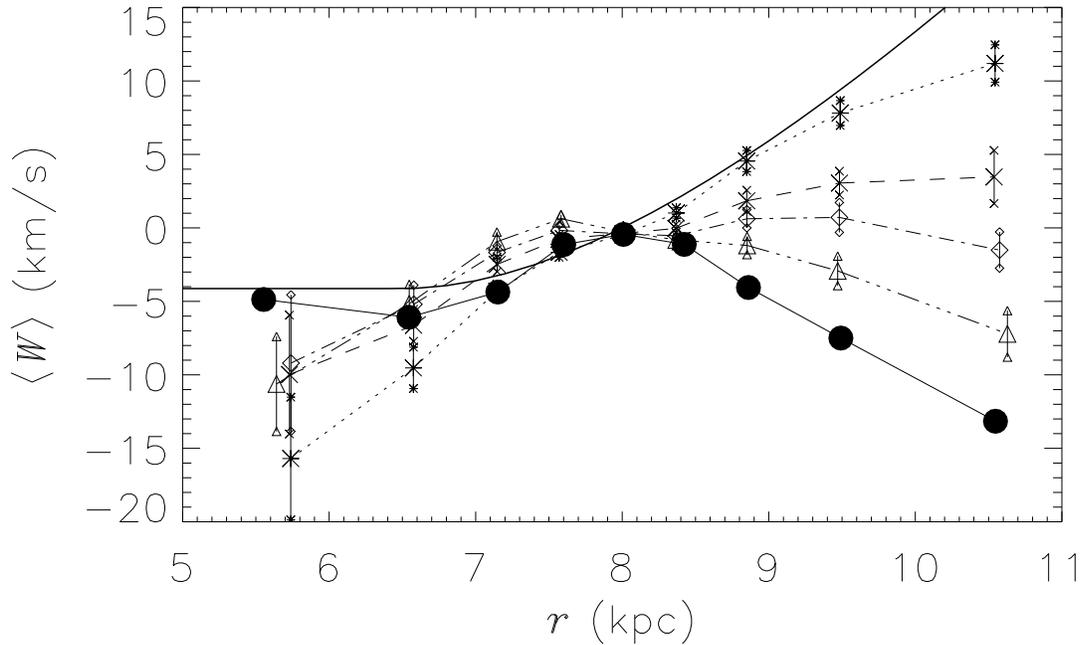}
\caption{The expected systematic vertical velocity from simulations
for four cases: a standard warp ($\ast$), a warp with half amplitude 
($\times$), a warp with half amplitude and a precession of $-13\kmskpc$ 
($\diamond$),  and a warp with half amplitude and a precession of 
$-25\kmskpc (\triangle)$.
In the first 3 cases the $\sigma_M$ is as given by Eq. (\ref{photerr}), 
while in the fourth model an additional 0.5 magnitude error is added.
Thirty simulated catalogues were generated for each case, and for each
catalogue median velocities were determined for each radial bin.  The
systematic velocity of each bin shown above represents the mean of the
30 median velocities, while the error bars are the standard deviation
of the 30 median velocities. 
The filled circles for the data and the solid line are the same as 
in Fig. \ref{vzrfig}, and are shown for comparison.
}
\label{h_allonone24s1}
\end{figure*}

To find the systematic vertical velocity $\overline{v}_z(r,\phi)$ 
due to a warp as described by Eq. (\ref{zwarp}), we consider 
Jeans continuity equation of stellar dynamics.
For the case of no radial motions we have
\begin{equation}
\frac{\partial \nu}{\partial t}
    + \frac{\partial (\nu \overline{v}_z)}{\partial z}
    + \frac{1}{r} \frac{\partial (\nu \overline{v}_\phi)}{\partial \phi}
= 0.
\label{continuity}
\end{equation}
If the stellar density $\nu$ is purely a function of $z'=z-Z_{\rm w}$ and
$\partial \overline{v}_z / \partial z = 0$,
the systematic vertical motion can be shown to be
\begin{equation}
\overline{v}_z(r,\phi) = 
\left( \frac{\overline{v}_\phi}{r} + \omega_{\rm p} \right) h(r) \cos \phi ~;
\label{sysvel}
\end{equation}
The systematic motion induced by a warp is simply the result
of the stars rotating with respect to the warp structure.
In our synthetic catalogues this systematic velocity is added to 
stars if they are in the warp (i.e. $r>r_{\rm w}$).

Fig. \ref{vzrfig} shows the systematic $\langle W \rangle$ 
for a typical synthetic catalogue 
generated with a non-precessing warp having the parameters obtained
from the sloped stellar distribution. 
This demonstrates that a non-precessing warp, consistent with the 
sloped distribution of our sample, should be easily detectable.
The $\langle W \rangle$
of the synthetic catalogue lies systematically below the theoretical 
curve primarily due to a bias introduced by the error in the 
photometric distances.

\section{Effect of bias, amplitude, and precession}

The bias introduced by distance errors can be understood after some
consideration. Stars whose distance is overestimated
are likely to have larger distance errors than those whose distance
are underestimated, as the distance error is normal in the
distance modulus. Together with the fact that in a magnitude limited
catalogue there are more stars closer to the Sun than further away, 
the skew in the distance errors means that 
stars with larger observed distances, $d_{\rm o}$, are more likely to be 
stars whose true distance, $d$, is smaller. 
The effect of the misplacement of a star to larger distances is that
its measured relative vertical motion $W_{\rm o}(d_{\rm o})$ will 
be smaller than its true relative vertical motion, $W(d)$. 
This can be seen from Eq. (\ref{zmotion}), for if differential 
Galactic rotation and vertical peculiar motions are neglected,
we can write $W(d)=( 4.74 \mu_b / \cos b) d + W_\odot$, and
similarly for $W_{\rm o}(d_{\rm o})$. Eliminating the common factor
$( 4.74 \mu_b / \cos b)$ it can be shown that 
$W_{\rm o}(d_{\rm o}) - W(d) = ( W(d) - W_\odot) (d_{\rm o}/d - 1)$.
Hence, as long as $W(d) < W_\odot$ and $d_{\rm o}/d > 1$, 
$W_{\rm o}(d_{\rm o})$ will be less than $W(d)$. 
As we look out from the Galactic center $r$ is proportional to $d$, so
if $d_{\rm o} > d$ due to error,  $r_{\rm o} > r$, and from the above 
$W_{\rm o}(r_{\rm o}) < W(r)$.
Finally, the kinematic signature of a warp increases outward, that is 
$W(r) < W(r_{\rm o})$, allowing us to write 
$W_{\rm o}(r_{\rm o}) < W(r_{\rm o})$;
the average relative systematic velocity measured at a given $r_{\rm o}$
is smaller than it's true systematic relative velocity at $r_{\rm o}$.

In addition to the above bias there are two other possible reasons why the 
observed systematic motions may be lower than that predicted for a 
long-lived warp. The first, trivially, is that
the actual amplitude of the warp may be smaller than we have estimated
from the observed slope of the spatial distribution. 
This may indeed be the case as radio
data yield a warp amplitude between .3 and .4\,kpc at a
galactic radius of 10\,kpc (\cite{BUR94}), while our warp parameters produce an
amplitude of .8\,kpc. In Fig. \ref{h_allonone24s1}
the signature expected from our standard warp ($r_{\rm h} = 15\kpc$) 
is shown together
with the signature of a warp with half amplitude ($r_{\rm h} = 30\kpc$).

Lastly, the warp may be precessing in the direction of Galactic
rotation.  Fig. \ref{h_allonone24s1} shows the expected signal of
a warp of half amplitude precessing at $-13\kmskpc$. While these
three effects, taken together, have significantly reduced the expected
velocity signature, we still have not recovered the negative
velocities seen in our Hipparcos sample. Could higher precession
rates, or higher errors in the photometric distances produce the
velocity signature that we detect? The fourth model in
Fig. \ref{h_allonone24s1} shows the effect of a precession rate of
$-25\kmskpc$ on a warp of half amplitude with elevated photometric 
error ($\sigma_M = 1. + {\rm max}(0,(m-6.5)/6) - M/12$).  
In this extreme case negative vertical motions are finally achieved.

\section{Discussion}

We have confirmed with a larger and more complete sample of distant OB
stars that the observed kinematics and structure
perpendicular to the Galactic plane are together inconsistent with the 
hypothesis of a long-lived non-precessing warp.  
However, with a revised model for the errors,
negative systematic velocities can be produced, but only
with excessively large precession rates, a smaller warp and large
photometric errors. A precession rate even as high as $-20\kmskpc$ is
questionable on physical grounds, for it would require that the warp
was precessing faster than Galactic rotation at a galactocentric
radius of only 11\,kpc!

Could the photometric error parameter even be as high as one magnitude? The
$\pi_{\rm P}$--$\pi_{\rm t}$ distribution of models generated with such errors
(see Fig. \ref{sigm_1}), compared to the observed $\pi_{\rm P}$--$\pi_{\rm t}$
distribution (Fig. \ref{pipi_nord}) would suggest not.

If we do not accept such large photometric errors and precession rates
we are left with the same conclusions that we arrived at in
our previous work; either the Galactic warp is not long-lived, or
there are other systematic motions present in the Galactic disk.
Galactic companions could be
responsible for a short-lived warp, or at least modify the velocities,
but the regularity of the Galactic warp out to at least 16\,kpc from
the Galactic center makes this seem unlikely. Nevertheless, numerical
simulations will have to be carried out to explore this possibility. 

Alternatively, the presence of other systematic
vertical oscillations in the Galactic disk could mask
the kinematic signature of the warp, but should also be
evidenced by accompanying deviations from the Galactic plane.  
The warp amplitudes as inferred from the OB stars found here
(.4 to .8\,kpc at $r=10\kpc$) are larger than those found in the HII gas 
(.3 to .4\,kpc at $r=10\kpc$, \cite{BUR94}) and the dust
(.3 kpc at $r=10\kpc$, \cite{FRE98}), and may 
suggest the presence of such structures. In addition 
Dehnen (1998) also finds that the kinematics of nearby 
stars are not consistent with a Galactic warp that begins within
the Solar Circle and could thus produce a local tilt in 
the stellar distribution. Indeed, the observed 
slopes and anomalous kinematic signature may not be due to a warp
at all, but to some other as yet unidentified phenomenon.

The possibility of other systematic motions being present in the 
disk complicates any evaluation of the kinematics of the Galactic warp. 
Until a deeper and more accurate astrometric survey is completed
the possibility of other kinematic effects cannot be excluded nor,
if present, be disentangled from the expected kinematic signature
of the warp in the stellar distribution. Such a survey 
will eventually be given by a future astrometric mission like GAIA,
which will also, via parallaxes, largely dispense with the bias introduced
by the photometric distance errors.

\section{Acknowledgments}
Thanks are extended to James Binney and Stefano Casertano for useful 
discussions, and the reviewer who contributed to the quality of this
contribution. R. Drimmel and R. L. Smart acknowledge support from 
the University of Turin. R. L. Smart also acknowledges support from 
the British Royal Society.


\end{document}